\documentclass[twocolumn,  superscriptaddress, pra]{revtex4}
\usepackage{amssymb}
\usepackage{amsmath}
\usepackage{graphicx}
\usepackage{dcolumn}
\usepackage{bm}
\usepackage{tabularx}

\begin{document}

\title{Generalized coherent-squeezed-state expansion for the quantum Rabi
model}
\author{Xiang-You Chen}
\affiliation{Department of Physics, Chongqing University, Chongqing 401330, China}
\author{Yu-Yu Zhang}
\email{yuyuzh@cqu.edu.cn}
\affiliation{Department of Physics, Chongqing University, Chongqing 401330, China}
\author{Hang Zheng}
\affiliation{Department of Physics and Astronomy, Shanghai Jiao tong University, Shanghai
200240, China}
\date{\today }

\begin{abstract}
We develop a systematic variational coherent-squeezed-state expansion for
the ground state of the quantum Rabi model, which includes an additional
squeezing effect with comparisons to previous coherent-state approach. For
finite large ratio between the atomic and field frequency, the essential feature
of the ground-state wave function in the super-radiant phase appears,
which has a structure of two delocalized wake packets. The single-peaked wave function
with one coherent-squeezed state works well even around the critical regime, exhibiting the
advantage over the coherent-state method. As the coupling increases to form strong correlations
physics in the vicinity of phase transition,
we develop an improved wave function with a structure of two Gaussian wave packets,
which is a linear superposition of two coherent-squeezed state. The
ground-state energy and the average photon number agree well with numerical
ones even in the strong-correlated regimes, exhibiting a substantial
improvement over the coherent-state expansion. The advantage of the
coherent-squeezed-state expansion lies in the inclusion of the second
coherent-squeezed state and the additional squeezed deformation of the wave
function, providing a useful tool for multi-modes spin-boson coupling
systems of greater complexity.
\end{abstract}

\date{\today }
\maketitle

\section{Introduction}

The quantum Rabi model describes the interaction of a two-level atom with a
single mode of the quantized electromagnetic field\cite{rabi}, which is the
simplest mode describe the interaction between spin systems and boson
systems. It has a wide application in cavity~\cite{baumann,nagy,Dimer} and
circuit~\cite{Wallraff,Niemczyk,pfd,fumiki} quantum electrodynamics (QED).
Recently, an emergence of quantum phase transition is found in the quantum
Rabi model when a ratio of the atomic transition frequency $\Delta$ to the
field frequency $\omega$ approaches infinity, $\Delta/\omega\rightarrow%
\infty $~\cite{plenio,liu,si}. This kind of complex behavior is challenged
to be captured by an accurate wave function when the coupling to the field
becomes sufficiently to form strong correlations physics.

Despite the fact that the quantum Rabi model has been solved exactly by a
Bargmann space technique~\cite{braak1,hengfan} and an coherent state method~%
\cite{chen}, respectively, where a numerical search for the zeros of
complicated transcendental functions is needed, an accurate expression of
the wave function remains elusive. Much works have developed an expansion of
the wave function of the field into a set of coherent states, such as a
generalized rotating-wave approximation (GRWA)~\cite{Irish}, a generalized
variational method~\cite{zheng,GangChen}, and adiabatic approximation~\cite%
{agarwal,Ashhab}. As previously studied, the oscillator state was considered
as a displaced state, and the squeezing effect of the oscillator part is
underestimated. With the consideration of the squeezing deformation of the
wave function, we have developed the generalized squeezing rotating-wave
approximation (GSRWA) with the coherent-squeezed state to solve the quantum
Rabi analytically, exhibiting an improvement over coherent state~\cite%
{zhang16,zhang17}. The physical role of the squeezing effect in the
coherent-squeezed state can not be overlooked in the intermediate coupling
regime. The purpose of the paper is to develop an alternative to a
reasonable wave function in the challenging regime of strong coupling to the
field. We shall focus on the problem of the strong-correlated ground state
of the quantum Rabi model, for which a super-radiant phase transition occurs in the limit $\Delta/\omega\rightarrow%
\infty $. A superposition of deforming polaron and
antipolaron has been used to explain the ground-state physics of the single-
and multi-modes spin-boson model~\cite{Shore,Ying,bera14}.

We develop a coherent-squeezed expansion to capture the strong-correlated
physics of the quantum Rabi model for finite large value of $\Delta/\omega$.
We first analyze the structure of the
displacements and squeezing deformations at strong coupling by exact
diagonalization, showing occurrence of two Gaussian wave packets for large value
of $\Delta/\omega$. The proliferation of the second Gaussian wave packet at
increasing coupling is associated with spin tunneling in the vicinity of the super-radiant phase transition. A second aspect of
our study will be to emphasize the improvement of the coherent-squeezed
state by including the squeezing deformations over previous coherent state especially in the strong-correlated
regime. The optimal displacements and squeezing parameters are determined
variationally. The validity of our improved coherent-squeezed expansion is
discussed by comparing with the coherent-state expansion as well as
numerical exact diagonalization, especially around the phase transition regimes.

The paper is organized in the following. We explore the structure of the
wave function in Sec. II by exact diagonalization. In Sec. III, we begin to
study the single-peak wave function with one coherent-squeezed state, and
compare with results obtained by one coherent state. Furthermore, we develop
the two coherent-squeezed expansion to study the ground state for large
value of $\Delta/\omega$. In Sec. V, we study the first-order phase
transition of the anisotropic Rabi model by the two coherent-squeezed state
with parity symmetry. We conclude the paper with a brief discussion.

\section{Deformed wavefunction}

The Hamiltonian of anisotropic Rabi model is
\begin{equation}  \label{H}
H=\frac{\Delta }{2}\sigma _{z}+\omega a^{\dagger }a+g(a^{\dagger }\sigma
_{-}+a\sigma _{+})+g\tau (a^{\dagger }\sigma _{+}+a\sigma _{-}),
\end{equation}%
where $\sigma _{i}(i=x,y,z)$\ are the Pauli matrices with the transition
frequency $\Delta $, $a^{\dagger }(a)$ is the creation (annihilation)
operator of photon with frequency $\omega $, and $g$ is the coupling
strength of rotating-wave (RW) interactions. In the paper $\omega $ is set
to be $1$. Here the parameter $\tau $ adjusts the relative weight between
the RW and counting-rotating-wave (CRW) interactions. The isotropic Rabi
model corresponds to $\tau =1$.

The quantum Rabi model has a discrete $Z_{2}$ symmetry associated with
parity operator $P=e^{i\pi N}$, where the excitation number $%
N=a^{\dagger}a+\sigma_{z}/2+1/2$ counts the total number of excitation quanta.
The parity operator satisfies $[H,P]=0$ and possesses two eigenvalues $p=\pm
1$ depending on whether the number of quanta is even or odd. In this paper,
we use the scaled coupling strength $\lambda =(1+\tau)g/\sqrt{\Delta \omega } $.
In the thermodynamic
limit $\Delta\rightarrow\infty$, there occurs the super-radiant phase
transition at the critical coupling strength $g_c=\sqrt{\Delta \omega }%
/(1+\tau)$~\cite{plenio,liu}.

\begin{figure}[tbp]
\includegraphics[scale=0.42]{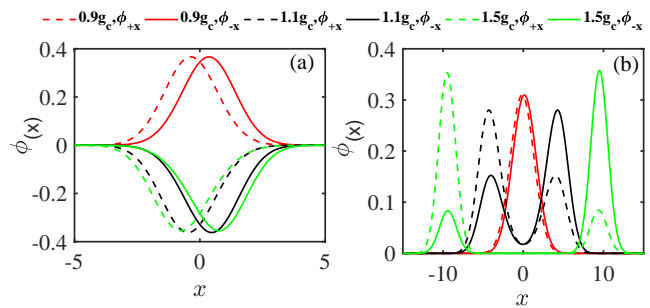}
\caption{Wave function in the ground state $\phi _{\pm x}$ in the projections of the spin part $|\pm x
\rangle $ as a function of the position $x$
 depending on the coupling strength $\protect\lambda=0.9, 1.1$ and
1.5 in the case of $\Delta/\protect\omega=1$ (a) and $\Delta/\protect\omega%
=100$ (b). }
\label{wave}
\end{figure}

It is interesting to explore the wave function in the
position representation, especially for the ground state in the vicinity of the quantum phase transition,
which can capture the deformation of the harmonic
oscillator. The ground state obtained form numerical
diagonalization is of the form
\begin{equation}
|\varphi \rangle =\sum_{n}^{N_{tr}}|n\rangle (c_{n+}|+x\rangle
+c_{n-}|-x\rangle ),
\end{equation}
where $N_{tr}$ is the truncated boson number in the Fock space, $c_{n\pm }$
are coefficients, and $|\pm x\rangle $ are the eigen-states of $\sigma _{x}$%
. In the position $x$ representation, the oscillator state $|n\rangle $ is
the usual harmonic oscillator wave function
\begin{equation}
\langle x|n\rangle =(\sqrt{\omega}/\sqrt{\pi}2^{n}n!)^{-1/2}e^{-\omega
x^{2}/2}H_{n}(\sqrt{\omega }x),
\end{equation}
where $H_{n}(\sqrt{\omega }x)$ are Hermite polynomials of order $n$.
Consequently, the wavefunction in the $x$ representation is
\begin{equation}
|\varphi \rangle =\phi _{+x}|+x\rangle +\phi _{-x}|-x\rangle ,
\end{equation}
with $\phi _{\pm x}=\sum_{n}c_{n\pm }\langle x|n\rangle $. In the case of $%
\Delta =0$, the oscillators displace in two directions by projecting onto $%
\left\vert \pm x\right\rangle $. It leads to doubly degenerate coherent
state for the displaced oscillators. For $\Delta\neq0$, a competition
between spin tunneling and oscillator displacement energy affect the
oscillator state $\phi _{\pm x}$. For $\Delta /\omega =1$ in Fig.~\ref{wave}
(a), as the scaled coupling $\lambda $ increases, the wave function $\phi
_{\pm x}$ in the $\left\vert \pm x\right\rangle $ projection always has a
single peak with a displacement. Extending the argument to large ratio $%
\Delta /\omega =100$, the single-peaked wave function stretches and then
splits into two as the coupling strength increases through the critical
value $\lambda _{c}=1$ in Fig.~\ref{wave}(b). On further increasing the
coupling $\lambda=1.5$, $\phi _{\pm x}$ moves away from each other and the
second peak disappears.

The key observation is the two-peaked wavefunction above $\lambda _{c}$ as
the ratio $\Delta /\omega$ tends infinity, which can play the role of a
thermodynamic limit and a super-radiant phase occurs~\cite{plenio,liu}.
Around the critical coupling $\lambda_c$, the wave functions
$\phi _{\pm x}$ become delocalized, which is a superposition of two Gaussian~\cite%
{Shore} or of two polarons~\cite{Ying,bera14}. It ascribes to the strong
competition between spin tunneling and oscillator displacement, which can
not be fulfilled by a single-peak wave function. The wave packets are consistent
with the delocalized wave function in the super-radiant phase of
the Dicke model~\cite{liu09,brandes}. The essential feature of the wave function
in the super-radiant phase in the infinite ratio $\Delta/\omega\rightarrow\infty$
appears for large ratio value, $\Delta/\omega=100$. For finite $\Delta/\omega$,
the system obeys the parity symmetry in the super-radiant phase and thus has
two lobes of the wave function, which is different from the discrepancy
in the infinite limit $\Delta/\omega\rightarrow\infty$.

Moreover, it is observed that the wave packets not only shift the displacement, but also have additional
squeezing deformations. However, in previous studies associated with coherent
states~\cite{Irish,GangChen}, the squeezing effects on the wave function
of the oscillators is overlooked. Recently, analytical solutions with an
coherent-squeezed state to the quantum Rabi model has been developed~\cite%
{zhang16,zhang17}, which exhibits an improvement over coherent-state approaches.
For finite large ratio of $\Delta/\omega$,
whether the coherent-squeezed state or
the coherent state can capture the wave functions around the critical region $\lambda=1$ remains unclear,
where the system becomes strong correlated.

\section{Single coherent-squeezed state ansatz}
For the scaled coupling strength $\lambda<\lambda_c$, the single-peaked wave function
works well in the ground state for the normal phase in Fig.~\ref{wave}(b).
In previous studies, coherent-state ansatz for the quantum Rabi model has
been employed to capture the oscillator state for small ratio of $%
\Delta/\omega$ and limiting coupling regimes~\cite{Irish,GangChen,zheng}.
As an improvement, we consider the additional squeezing effects of the
oscillator by comparing with the coherent state. The single-peaked wave function can be expressed by one
coherent-squeezed state ($1$CSS) as
\begin{equation}  \label{css}
\left\vert \psi _{1CSS}^{+}\right\rangle = C_{1} \left\vert
+x\right\rangle\otimes \left\vert f^{(1)}\right\rangle + C_{2} \left\vert
-x\right\rangle \otimes \left\vert f^{(2)}\right\rangle,
\end{equation}%
where the oscillator state is the coherent-squeezed state
\begin{equation}
\left\vert f^{(k)}\right\rangle =U^{\dagger}(\beta_{k})S^{\dagger }(\xi
_{k})\left\vert 0\right\rangle.
\end{equation}
Here, the coherent operator is $U(\beta _{k})=e^{\beta_{k}(a^{\dagger }-a)}$
with the variational displacements $\beta _{k}$, and the squeezing operator
is $S(\xi_{k})$ =$e^{\xi_{k}(a^{2}-a^{\dagger ^{2}})}$ with the squeezing
parameters $\xi _{k}$. Since the ground state is associated with the even
parity, satisfying $P\left\vert \psi _{1CSS}\right\rangle =\left\vert
\psi_{1CSS}\right\rangle$. The variational parameters are constrained as $%
C_1=-C_2=1/\sqrt{2}$, $\beta_1=-\beta_2=\beta$, and $\xi_1=\xi_2=\xi$. This more flexible ansatz facilitates us to obtain
the variables $\beta$ and $\xi$. Especially, by setting the squeezing $\xi=0$%
, the one coherent-squeezed state $\left\vert \psi _{1CSS}\right\rangle$
reduces to one coherent state ($1$CS) $\left\vert \psi _{1CS}\right\rangle$ with the
coherent state $\left\vert f^{(k)}\right\rangle =U^{\dagger
}(\beta_{k})\left\vert 0\right\rangle$.

The Hamiltonian of the isotropic and anisotropic quantum Rabi model is rewritten as
\begin{equation}
H=\frac{\Delta }{2}\sigma _{z}+\omega a^{\dagger }a+\alpha (a^{\dagger
}+a)\sigma _{x}+\gamma (a^{\dagger }-a)i\sigma _{y}.
\end{equation}
where parameters $\alpha =g(1+\tau)/2$ and $\gamma =g(\tau-1)/2$.
With the one coherent-squeezed state $\left\vert \psi _{1CSS}\right\rangle$ in Eq.(\ref{css}),
the corresponding energy is given by
\begin{eqnarray}
E_{g}^{1CSS} &=&\omega (\sinh ^{2}2\xi +\beta ^{2})-2\beta \alpha  \notag
\label{gsenergy} \\
&&-(\frac{\Delta }{2}+2\gamma \beta \eta^2 )e^{-2\beta ^{2}\eta ^{2}},
\end{eqnarray}%
where $\eta =\mathtt{cosh}(2\xi )-\mathtt{sinh}(2\xi )$. The variational
energy is minimized according to $\partial E_{1g}/\partial \beta =0$ and $%
\partial E_{1g}/\partial \xi =0$, respectively. Hence, the displacement $%
\beta $ and squeezing parameter $\xi $ are variationally determined. One
obtain the equations for the isotropic case $\tau =1$
\begin{equation}
\omega (e^{4\xi }-e^{-4\xi })-4\Delta \beta ^{2}e^{-4\xi }e^{-2\beta
^{2}\eta ^{2}}=0,
\end{equation}%
\begin{equation}
(\omega \beta -g)+\Delta \beta e^{-4\xi }e^{-2\beta ^{2}\eta ^{2}}=0.
\end{equation}
In the limit $\Delta /\omega \rightarrow \infty $,
the values of $\beta $ and $\xi $ are given approximately
\begin{equation}
\beta \simeq \frac{g}{\Delta},
\end{equation}%
\ and%
\begin{equation}
\xi =\frac{1}{8}\ln (1+\frac{4g^{2}}{\omega \Delta }).
\end{equation}%
We observe that the squeezing influence $\xi $ plays a more crucial role
as the coupling strength approaches the critical value $\lambda_{c}=1$,
for which the atom becomes strongly correlated with the oscillator.

Using the single-peaked wave function $\left\vert\psi _{1CSS}\right\rangle$,
the mean photon number can be expressed as
\begin{equation}
\left\langle a^{\dagger }a\right\rangle =\sinh ^{2}2\xi+\beta^{2}.
\end{equation}

For large ratio of $\Delta/\omega=100$, the one coherent-squeezed state $\left\vert \psi _{1CSS}\right\rangle$ works
relatively well for the ground state by compared with one coherent state $%
\left\vert \psi _{1CS}\right\rangle$ in Fig.~\ref{1css}. The ground-state energy $E_{g}$ and the mean photon
number for the isotropic $\tau=1$ and anisotropic $\tau=1.5$ cases
agree well with numerics for the scaled coupling $\lambda<1$, but the
results of the single coherent state $\left\vert \psi _{1CS}\right\rangle$
get worse as $\lambda$ increases to the critical value $1$. The success of the
coherent-squeezed state comes from the squeezing effects especially as the
coupling increases to the strong-correlated regime. However, the single-peak
state $\left\vert \psi _{1CSS}\right\rangle$ presents deficiencies for $%
\lambda>1$, where the competition between the spin tunneling and the
oscillator displacement becomes strong.

\section{Two coherent-squeezed state ansatz}

\begin{figure}[tbp]
\includegraphics[scale=0.4]{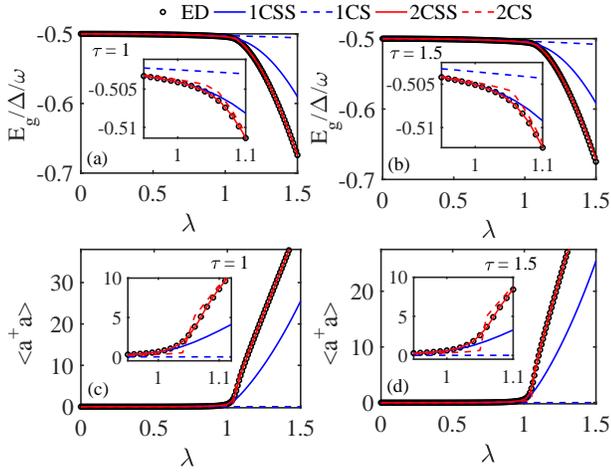}
\caption{Ground-state energy $E_g/(\Delta\protect\omega)$ and mean photon
number $\left\langle a^{\dagger }a\right\rangle$ for the single
coherent-squeezed state (1CSS) (blue solid line) and two coheren-squeezed
state (2CSS) (red solid line) as a function of the rescaled coupling
strength $\protect\lambda$ for the isotropic $\protect\tau=1$ (a)(c) and
anisotropic $\protect\tau=1.5$ Rabi model (b)(d) with the large detuning $%
\Delta/\protect\omega=100$. The results obtained by the singel coherent
state (1CS) (blue dashed line) and two coherent state (2CS) (red dashed
line) are listed for comparison. The inset shows the results from $\protect%
\lambda=0.9$ to $1.1$. }
\label{1css}
\end{figure}
As the coupling strength exceeds the critical value $\lambda>\lambda_c$,
it becomes a strongly-correlated system and two Gaussian of the wave function
emerges in Fig.~\ref{wave} (b). As an improvement, a two-peaked
state $\left\vert \psi _{2CSS}\right\rangle$ is proposed by a
linear combination of two coherent-squeezed state ($2$CSS)
\begin{eqnarray}  \label{state2}
&&\left\vert \psi _{2CSS}^{+}\right\rangle=\left\vert +x\right\rangle
\otimes(C_1\left\vert +f^{(1)}\right\rangle+C_2\left\vert
+f^{(2)}\right\rangle)  \notag \\
&&-\left\vert -x\right\rangle \otimes(C_1\left\vert
-f^{(1)}\right\rangle+C_2\left\vert -f^{(2)}\right\rangle ),
\end{eqnarray}
where the coherent-squeezed state $\left\vert \pm
f^{(k)}\right\rangle=e^{\mp\beta_{k}(a^{\dagger
}-a)}e^{\xi(a^{2}-a^{\dagger ^{2}})}|0\rangle$. The symmetry of the even
parity enforces the chosen relative sign between the $\pm x$ components of
the state in Eq. (\ref{state2}). Since the parity breaks at the critical point
in the limit $\Delta/\omega\rightarrow\infty$, there the ground state remains
in the even parity due to finite large value of $\Delta/\omega=100$.
Variables $\beta_{1(2)}$ and $\xi$ in the coherent-squeezed states are
taken as free parameters and can be determined by minimizing the energy
$E^{2CSS}_{g}=\left\langle\psi _{2CSS}^{\dagger}\right\vert
H\left\vert \psi_{2CSS}^{\dagger}\right\rangle$ in the Appendix. Meanwhile, the
two-coherent state ($2$CS) $\left\vert \psi _{2CS}\right\rangle$ can be given by the
coherent state $\left\vert \pm
f^{(k)}\right\rangle=e^{\mp\beta_{k}(a^{\dagger }-a)}|0\rangle$ instead of the
coherent-squeezed state in Eq.(\ref{state2}).

Fig.~\ref{1css} shows the $\texttt{2CSS}$ ansatz including the addition of a second coherent-squeezed state
dramatically improves the ground-state energy and mean photons,
which are in very good agreement with numerical results both for the isotropic and anisotropic case. Since the two-peaked
wave function $\left\vert \psi _{2CSS}\right\rangle$ correctly predicts an enhancement of the spin
tunneling around the critical value $\lambda=1$ in comparison to the single-peaked state. However, the
results of two coherent state $\left\vert \psi _{2CS}\right\rangle$ remains
a deviation in the vicinity of the critical coupling strength. The failure of the coherent state
attributes to the ignorance of the squeezing variance of the wave functions.

\begin{figure}[tbp]
\includegraphics[scale=0.45]{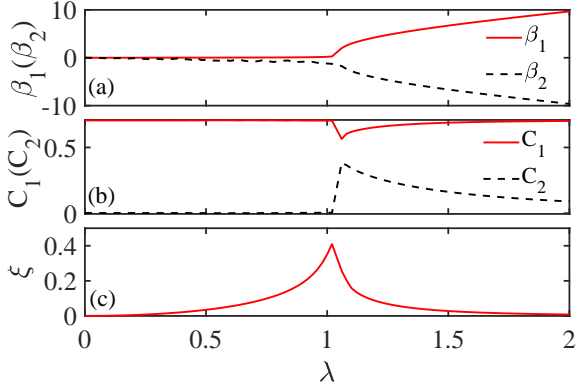}
\caption{ Displacement Variables $\protect\beta_1$ (red solid line) and $%
\protect\beta_2$ (black dashed line) (a), the coefficients $C_1$ (red solid
line) and $C_2$ (black dashed line) (b), and the squeezing $\protect\xi$ (c)
for the two coherent-squeezed state (2CSS) for the isotropic Rabi model $%
\protect\tau=1$.}
\label{beta}
\end{figure}

The corresponding variables $\beta_{1(2)}$, $C_{1(2)}$ and squeezing parameter $\xi$ are calculated by
minimizing the ground-state energy $E^{2CSS}_{g}$ in Fig.~\ref{beta}. In the weak coupling regime $\lambda<1$,
coefficient $C_1$ approaches to $1$ and $C_2$ tends to $0$, which
corresponds to single-peaked wave function. As coupling exceeds the critical value $1$, $C_2$
increases dramatically and the additional second coherent-squeezed state plays an important effect.
It corresponds to the occurrence of the second peak of the delocalized wave function for $\lambda>1$ in Fig.~\ref{wave}(b).
Then, the probability of the second peak decreases with further increasement of $%
\lambda$, which agrees well with vanishing of the second peaks for $%
\lambda=1.5$ in Fig.~\ref{wave}(b). Meanwhile, the displacements of two
coherent-squeezed state $\beta_1$ and $\beta_2$ become larger as the
coupling strength exceeds the critical value $\lambda>1$. It reveals that
two Gaussian wave packets move in the opposite directions in
Fig.~\ref{wave}(b). Moreover, the squeezing parameter $\xi$ increases and
reaches the maximum value at the critical value $\lambda=1$. It means that
the squeezing effect of the oscillator state can not be overlooked
especially in the critical regime. This motivates the inclusion of
additional coherent-squeezed state in the ground-state ansatz, which further
captures the spin tunneling in the strong-correlated regimes and squeezing influence of the wave functions.

\section{First-order quantum phase transition in the anisotropic Rabi model}
Besides the second-order super-radiant phase transition, a additional first-order phase transition occurs at the critical value $g^{(1)}_c=%
\sqrt{\Delta\omega/(1-\tau^2)}$ in the anisotropic Rabi model~\cite{hengfan,si}. When the atom-photon RW coupling strength
becomes stronger than that of the CRW terms, e.g. $\tau<1$, there exists energy-level crossing
between the ground state and the first-excited state. A natural question follows as to
the ground-state wave function in the first-order phase transition of the anisotropic quantum Rabi model.
As the coupling strength exceeds the critical value $g_c^{(1)}$, the even parity symmetry of the ground
state breaks down and the first-excited state with the odd parity becomes to the
lowest-energy state. It is reasonable to describe the first-excited state by the two coherent-squeezed
state
\begin{eqnarray}\label{odd}
&&\left\vert \psi_{2CSS}^-\right\rangle = \left\vert +x\right\rangle
\otimes\lbrack C_{1}\left\vert +f^{(1)}\right\rangle +C_{2}\left\vert
+f^{(2)}\right\rangle ]  \notag \\
&&+\left\vert-x\right\rangle \otimes \lbrack C_{1}\left\vert
-f^{(1)}\right\rangle +C_{2}\left\vert -f^{(2)}\right\rangle],
\end{eqnarray}
where the coefficients and the coherent-squeezed state $\left\vert \pm f\right\rangle$
for the spin states $|\pm x\rangle$ satisfy the odd parity.
With its comparison to the ground state $\left\vert \psi_{2CSS}^{+}\right\rangle$ (\ref{state2})
with the even parity symmetry, $\left\vert \psi_{2CSS}^-\right\rangle$
changes the sign of the wave function for the $|-x\rangle$ part due to the odd parity.

Using two coherent-squeezed state $\left\vert \psi_{2CSS}^{\pm}\right\rangle$%
, the ground-state and first-excited-state energy can be obtained as $%
E^{\pm}=\left\langle\psi_{2CSS}^{\pm}\right\vert H^{Rabi}\left\vert
\psi_{2CSS}^{\pm}\right\rangle$ (see Appendix). The variational
displacement, squeezing and coefficients can be determined by minimizing the
energy $E_{g}^{\pm}$.

\begin{figure}[tbp]
\includegraphics[scale=0.42]{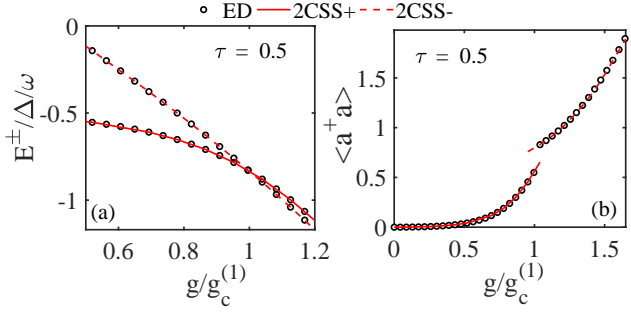}
\caption{(a) Energy levels $E^{\pm}$ for the ground state $|\psi _{2CSS}^{+}\rangle$
(red solid line) and the first-excited state $|\psi _{2CSS}^{-}\rangle
$ (red dashed line) as a function of the scaled coupling strength $g/g_c^{(1)}$ for
the anisotropic Rabi $\protect\tau=0.5$.
Numerical results by the exact diagonalization are listed (black circle).
(b) Mean photon number $a^{\dagger}a$ in the ground state as a function of $g/g_c^{(1)}$.}
\label{eg2}
\end{figure}
Energy levels $E^{\pm}$ for the ground state $\left\vert
\psi_{2CSS}^{+}\right\rangle$ and the first-excited state $\left\vert
\psi_{2CSS}^{-}\right\rangle$ agree well with the numerical ones in Fig.~\ref%
{eg2}(a). Energy-levels crossing is captured at the critical value $%
g_c^{(1)}$ by two coherent-squeezed state, exhibiting the accuracy of our
methods. Moreover, the average photons $\langle a^{\dagger}a\rangle$ in the ground state
shows a discontinuous transition in Fig.~\ref{eg2}(b), exhibiting the first-order phase transition.
The success of our method lies in the inclusion of the second coherent-squeezed state and the additional
squeezing effect of the wave functions.

\section{Conclusions}

We have developed a coherent-squeezed-state expansion for the isotropic and
anisotropic quantum Rabi model, which includes the squeezing effect of the
oscillator state and shows an improvement over previous coherent-state
methods. The improved coherent-squeezed state allows squeezing distortion of the wave packets
as well as displacement.
Excellent agreement is found with numerical ones. We have
constructed a simple physical picture in terms of displacements and
squeezing influence of the wave function in the vicinity of the phase
transition, which turns out to be validity of coherent-squeezed-state expansion.

\section{Acknowledgments}

This work was supported in part by the Natural Science Foundation of China
under Grant No.11847301 and by the Fundamental Research Funds for the
Central Universities under Grant No. 2019CDXYWL0029 and No.2019CDJDWL0005.

\appendix

\section{Energy of the two-coherent-squeezed state}

Using the two coherent-squeezed state with even parity $\left\vert \psi_{2CSS}^{+}\right\rangle$ in Eq. (\ref{state2}),
the energy can be written as $E^{+}=E^{atom}+E^{ph}+E^{iso-int}+E^{ani-int}$, consisting of
\begin{widetext}
\begin{eqnarray}
E^{atom} &=&\left\langle \psi _{2CSS}|\frac{\Delta }{2}\sigma _{z}|\psi
_{2CSS}\right\rangle   \notag \\
&=&-\Delta \lbrack C_{1}^{2}\left\langle +f^{(1)}|-f^{(1)}\right\rangle
+2C_{1}C_{2}\left\langle +f^{(1)}|-f^{(2)}\right\rangle
+C_{2}^{2}\left\langle +f^{(2)}|-f^{(2)}\right\rangle ], \\
E^{ph} &=&\left\langle \psi _{2CSS}|\omega a^{\dagger }a|\psi
_{2CSS}\right\rangle   \notag \\
&=&\omega \lbrack 2C_{1}^{2}\beta _{1}^{2}+2C_{2}^{2}\beta
_{2}^{2}+2C_{1}C_{2}\beta _{1}\beta _{2}(\left\langle
+f^{(1)}|+f^{(2)}\right\rangle +\left\langle -f^{(1)}|-f^{(2)}\right\rangle
)]  \notag \\
&&+\omega \sinh ^{2}2\lambda \{2C_{1}^{2}+2C_{2}^{2}+C_{1}^{2}e^{-2\beta
_{1}^{2}\eta ^{2}}+C_{2}^{2}e^{-2\beta _{2}^{2}\eta ^{2}}+  \notag \\
&&2C_{1}C_{2}[1-\eta ^{2}(\beta _{1}-\beta _{2})^{2}](\left\langle
+f^{(1)}|+f^{(2)}\right\rangle +\left\langle -f^{(1)}|-f^{(2)}\right\rangle
)]  \notag \\
&&+\omega \sinh 2\lambda \lbrack 2C_{1}C_{2}\eta (\beta _{1}-\beta
_{2})^{2}(\left\langle +f^{(1)}|+f^{(2)}\right\rangle +\left\langle
-f^{(1)}|-f^{(2)}\right\rangle )], \\
E^{iso-int} &=&\left\langle \psi _{2CSS}|\alpha (a^{\dagger }+a)\sigma
_{x}|\psi _{2CSS}\right\rangle   \notag \\
&=&-2\alpha \lbrack C_{1}^{2}\beta _{1}+C_{2}^{2}\beta _{2}+C_{1}C_{2}(\beta
_{1}+\beta _{2})(\left\langle +f^{(1)}|+f^{(2)}\right\rangle +\left\langle
-f^{(1)}|-f^{(2)}\right\rangle ), \\
E^{ani-int} &=&\left\langle \psi _{2CSS}|\gamma (a^{\dagger }-a)i\sigma
_{y}|\psi _{2CSS}\right\rangle   \notag \\
&=&-4\gamma \eta ^{2}[C_{1}^{2}\beta _{1}\left\langle
+f^{(1)}|-f^{(1)}\right\rangle +C_{1}C_{2}\left\langle
+f^{(1)}|-f^{(2)}\right\rangle (\beta _{1}+\beta _{2})+C_{2}^{2}\left\langle
\beta _{2}+f^{(2)}|-f^{(2)}\right\rangle ].
\end{eqnarray}
\end{widetext}
where the overlap of the coherent-squeezed state is
\begin{equation}
\left\langle +f^{(k)}|\pm f^{(k^{\prime })}\right\rangle =\left\langle -f^{(k)}|\mp f^{(k^{\prime })}\right\rangle=e^{-\eta ^{2}\frac{(\beta _{k}\mp \beta_{k^{\prime }})^{2}}{2}}.
\end{equation}

For the two coherent-squeezed state with the odd parity $\left\vert \psi_{2CSS}^{-}\right\rangle$ in Eq. (\ref{odd}),
the energy is similarly obtained as $E^{-}=-E^{atom}+E^{ph}+E^{iso-int}-E^{ani-int}$.

\end{document}